\documentclass{article}
\usepackage[T1]{fontenc} 
\usepackage[utf8]{inputenc} 
\usepackage{ismir,amsmath,cite,url,amssymb}
\usepackage{graphicx}
\usepackage{color}
\usepackage{multirow}

\usepackage{lineno}

\title{DDX7: Differentiable FM Synthesis of Musical Instrument Sounds}

\oneauthor
{Franco Caspe, Andrew McPherson, Mark Sandler}
{Queen Mary University of London \\ {\tt \{f.s.caspe, a.mcpherson, mark.sandler\}@qmul.ac.uk}}

\def\authorname{F. Caspe, A. McPherson, and M. Sandler}

\usepackage[bookmarks=false,pdfauthor={\authorname},pdfsubject={\papersubject},hidelinks]{hyperref}

\begin{document}

\maketitle
\begin{abstract}
FM Synthesis is a well-known algorithm used to generate complex timbre from a compact set of design primitives. Typically featuring a MIDI interface, it is usually impractical to control it from an audio source.
On the other hand, Differentiable Digital Signal Processing (DDSP) has enabled nuanced audio rendering by Deep Neural Networks (DNNs) that learn to control differentiable synthesis layers from arbitrary sound inputs. The training process involves a corpus of audio for supervision, and spectral reconstruction loss functions. Such functions, while being great to match spectral amplitudes, present a lack of pitch direction which can hinder the joint optimization of the parameters of FM synthesizers.
In this paper, we take steps towards enabling continuous control of a well-established FM synthesis architecture from an audio input. Firstly, we discuss a set of design constraints that ease spectral optimization of a differentiable FM synthesizer via a standard reconstruction loss. Next, we present Differentiable DX7 (DDX7), a lightweight architecture for neural FM resynthesis of musical instrument sounds in terms of a compact set of parameters. We train the model on instrument samples extracted from the URMP dataset, and quantitatively demonstrate its comparable audio quality against selected benchmarks.

\end{abstract}

\section{Introduction}\label{sec:introduction}


Sound generation and transformation tools are ubiquitous in music composition and production. The electronic synthesizer has enabled musicians to access forms of timbre beyond the capabilities of acoustic or amplified instruments. 
Chowning's FM Synthesis \cite{chowning1973synthesis} is a widely used technique that is flexible to create complex, harmonic and inharmonic spectra from a reduced set of controls. Its long-standing presence in the audio industry has shaped traditional and contemporary sound design techniques \cite{lavengood2019makes,miranda2012computer,stevens2021teaching} across musicians and producers, with a wide number of current commercial synthesizer keyboards, modules and software plugins featuring it.


For most of its history, the digital synthesizer has been inextricably associated with the MIDI keyboard. Several historical technical and social factors contributed to making the keyboard the de-facto control interface for the synthesizer \cite{pinch1998social}. The legacy of this association continues to steer synth design, favouring triggers and envelope generators over continuous control strategies.

Developing alternative interfaces for controlling digital synthesis remains an active area of research \cite{the_nime_reader}. Many such digital musical instruments (DMIs) are based on mappings between sensor data and synthesis parameters \cite{west2021making,regimbal2021interpolating}.
An alternative approach uses features extracted from the audio signal of an acoustic instrument to control a digital synthesis process.
In that context, previous works employed audio signals from musical instrument as oscillators \cite{asdss_paper}, in a configuration that can be seen as a special application of Adaptive Digital Audio Effects \cite{1678000,nuimeprn4166}.

More recently, Neural Audio Synthesis (NAS) algorithms have employed Deep Neural Networks (DNNs) to map audio features
to synthesizer parameters, in tasks such as realistic audio synthesis from a compact set of
control signals\cite{engel_ddsp_2020,hayes2021neural,michelashvili2020hierarchical}, timbre transfer from one instrument to another \cite{cifka2021self,50554} and enhancement of symbolic musical expression \cite{wu2021midi}.
While the results are impressive, we observe that the employed synthesis architectures
are overly complex, featuring dozens of time-evolving parameters that hinder intervention, with users defaulting to indirect methods to intervene with the synthesis process, such as network bending \cite{yee2021studio}.


We are interested in an audio-based, continuous control technique for a well-established synthesis architecture, that can potentially offer musicians similar sound design primitives and outcomes available to keyboard players. Furthermore, we want the model to be compatible with live use, therefore it should be able to work in real-time.


In this work, we take steps towards enabling interpretable sound design controls for NAS algorithms, and present Differentiable DX7 (DDX7), a causal and lightweight DNN architecture that maps continuous audio features to the synthesis parameters of a well-known FM synthesizer.
We train DDX7 in single instrument datasets and evaluate its resynthesis performance against selected benchmarks. We provide full source code, and an online supplement\footnote{\url{https://fcaspe.github.io/ddx7}} with audio examples and a preliminary analysis of the model's real-time execution capabilities.








\section{Background}\label{sec:background}
\subsection{Linear FM Synthesis}


Linear FM modulation for audio signals, originally described by Chowning \cite{chowning1973synthesis} in the early 1970s, is a well-known sound design technique that powered the first massively available and commercially successful digital synthesizer, the Yamaha DX7, defining an era in music production \cite{lavengood2019makes}. Since then, it that has retained a fair amount of attention by the research community \cite{nielsen2020practical}, being present in a variety of topics from timbre semantic analysis \cite{hayes2021neural}, to empirical \cite{stevens2021teaching} and computer assisted sound design \cite{masuda2021quality}, and adaptive effects \cite{nuimeprn4166}.
Commercially, manufacturers continue to release as of today a sizable number of new instruments based on this technology \cite{fm_devices_1,fm_devices_2}.

Linear FM synthesis is actually a phase modulation technique aimed at generating different timbres with few parameters and low computational resources. Expressed in term of sine waves, the instantaneous FM modulated signal can be written as shown in Equation \ref{eq:fmmod}, where $f_c$ is the carrier frequency, $f_m$ is the modulator frequency and $I$ is the modulation index. In this work, we denote this particular linear phase modulation technique simply as FM Synthesis.
\begin{equation}
    y(t) = sin(2 \pi f_c t + I sin(2 \pi f_m t))
\label{eq:fmmod}\end{equation}

The side-bands of an FM signal are equally spaced around the carrier frequency, and their separation determined by $f_m$, while the number of harmonics depends on the modulation index and follows a Bessel function of the first order (Equation \ref{eq:bessel}).

\begin{equation}
    y(t) = \sum_{n=-\infty}^{n=+\infty} J_n(I) sin( 2 \pi( f_c + n f_m)\cdot t)
\label{eq:bessel}
\end{equation}

By controlling the modulation indexes and ratios, users can generate rich and complex timbre with a small set of oscillators modulated in frequency. For instance, the simple modulator-carrier setup can be extended with a new carrier that is modulated by the output of the previous pair. This \textit{stacked} arrangement spreads the bands of the initial modulator-carrier pair across another carrier, adding a new layer of complexity to the timbre. Furthermore, outputs from many stacks can be mixed in an \textit{additive} approach. Next, by setting the frequency ratio $r$ in such a way that $f_c = r \cdot f_m$, with $r \in \mathbb{Q}$, FM generates harmonic spectra.




\subsection{The DX7 Synthesizer}

The Yamaha DX7 synthesizer is probably the best-known FM synthesizer, and it features a linear FM synthesis architecture that has been previously used by other synth models as well, making it an excellent candidate for integration to a continuous control strategy.

The sound of the DX7 is generated by six frequency-modulated sinusoidal oscillators, and it is programmed by means of setting up a \textit{patch}.
For each oscillator, a patch describes their routing (i.e. how the oscillators are interconnected in a stacked or additive fashion), the Attack-Decay-Sustain-Release (ADSR) envelope generator parameters and the frequency ratios with respect to the note that is being played. Fixed frequencies can also be assigned to the oscillators.
The ADSR parameters control the \textit{output levels} of each oscillator; affecting their output volume or modulation index depending on their interconnection.
Finally, in the DX7, the oscillators' frequency ratios and routing remain \textit{fixed} during audio rendering. Sound dynamics are generated mainly by the ADSR envelopes controlling the volume or modulation index of the carriers and modulators respectively.

\subsection{Sound Matching}

Sound matching is the process of estimating a set of synthesis parameters that approximate a given audio signal as best as possible.
The task of matching an audio input to synthesizer controls is not new especially for FM, where evolutionary methods have been used to optimize a patch for a particular given sound \cite{horner1993machine,masuda2021quality}.

More recently, DNNs have been studied for supervised sound matching from an annotated dataset of patches and synthesized audio excerpts. The approaches include classification \cite{yee-king_automatic_2018}, where the network predicts a patch based on an input spectrogram, variational inference \cite{le2021improving}, with the DNN learns an invertible mapping between a dataset of audio and patches, and multi-modal analysis \cite{chen2022sound2synth} that employs multiple aggregated audio features for prediction.

While these methods are effective, they usually process long audio windows on the order of seconds, required to estimate the values of the ADSR envelope generator. Furthermore they can only estimate parameters from audio excerpts at specific pitch values. 
These approaches are not suitable for continuous control of a synthesizer from an audio signal.


\subsection{Neural Audio Synthesis}

DNNs can capture complex relations from a dataset. This can be exploited in a generative approach to produce realistic sounding audio. Neural Audio Synthesis (NAS) algorithms employ DNNs as powerful synthesizers that can capture structure and nuances from a corpus of music and produce new audio with similar characteristics during inference. Furthermore, these algorithms can be conditioned at train or inference time, modifying the output on-the-fly.

Since the introduction of the seminal autoregressive architectures Wavenet \cite{oord2016wavenet} and SampleRNN \cite{sample_rnn}, the NAS field has included generation approaches such as Generative Adversarial Neural Networks (GANs) \cite{nistal_drumgan_2020,engel_gansynth_2019,lavault2022stylewavegan}, probabilistic models \cite{esling_flow_2020,caillon2021rave} and style transfer methods \cite{cifka2021self,huang_timbretron_2019}.
Furthermore, a branch of the field has focused on controllable music generation, in an effort to bring interaction possibilities to users, with NAS algorithms supporting control inputs such as MIDI \cite{wu2021midi}, timbre descriptors \cite{nistal_drumgan_2020} and pitch and loudness signals \cite{michelashvili2020hierarchical,engel_ddsp_2020,hayes2021neural}.


\subsection{Differential Digital Signal Processing}

Differential Digital Signal Processing (DDSP)\cite{engel_ddsp_2020} is one approach for efficient, high quality audio rendering from a compact set of input controls. 
It biases a DNN towards generating and processing audio by the insertion of signal processing elements, such as oscillators and filters, in the network structure. 
These are implemented using differentiable operators from a neural network training framework and therefore can back-propagate gradients during training.


DDSP methods have been successfully employed in resynthesis and tone transfer\cite{50554} tasks, where they are conditioned on a set of time-evolving inputs, including the fundamental frequency of the target signal, and generate audio on a frame-by-frame basis.
Some approaches such as the Neural Source Filter (NSF) \cite{michelashvili2020hierarchical,wang2019neural} or the Neural Waveshaper \cite{hayes2021neural} learn to control a non-linear filter that shapes a harmonic source towards a particular target sound. 
Other DDSP architectures \cite{engel_ddsp_2020,wu2021midi} directly drive a Harmonic plus Noise (HpN) synthesizer \cite{stylianou1995high}, effectively learning a mapping between the input controls and the synthesizer parameters that generate the output.

While the previous resynthesis architectures can generate realistic tone transfer, there is little users can do to manipulate the resulting audio other than controlling the pitch and loudness inputs. The DNN controls spectral modelling algorithms, which do not have musically meaningful parameters. In this work, we present a differentiable FM synthesizer module that features a compact set of well-known sound design controls driven by a DNN, enabling a potential user intervention into the synthesis process.

While the idea of differentiable FM is not new, we have not been able to find published literature with details and evaluations of systems employing it.
We highlight two web repositories, one involving a 2-oscillator FM optimization strategy of audio excerpts\cite{andreas_jansson_repo} and another presenting an extension of the original DDSP project with a Differentiable FM synthesizer \cite{juan_alonso_repo}, where the experiments fail to reproduce musical instrument sounds.
Our approach imposes a set of constraints on the FM synth that allows a DNN to generate instrument sounds.





\subsection{Training objective for DDSP resynthesis}

The training process for the DDSP algorithms usually involves a multi-scale spectral (MSS) loss function that includes the $L_1$ distance of the amplitude spectrograms of a synthesized and a target audio excerpt, in linear and logarithmic form\cite{engel_ddsp_2020}. This function is used as a reconstruction loss, with the DDSP model aiming to replicate the target spectra during training.




Despite its widespread use, the MSS loss and more generally, spectral-based distance metrics present pitch-based failure modes that can conspire against the generalization capabilities of NAS algorithms when trained with gradient descent, as demonstrated by Turian et. al. \cite{turian_im_2020}. The authors show how such functions fail to propagate informative gradients for fine-tuning oscillator frequencies towards a frequency target due to fine grained ripple on the loss surface. Furthermore, they indicate that for the MSS distance, jointly optimizing amplitude and frequency generates misleading gradients for both tasks; this loss function can match spectral amplitudes only when the harmonics of target and prediction are aligned in frequency.

We argue that the MSS loss works well for DDSP because of two main reasons: Firstly, these models drive highly parameterized spectral modelling synthesizers with fine control on the output, either in the form of filter-distortion DNNs \cite{michelashvili2020hierarchical,wang2019neural}, multiple parallel waveshapers \cite{hayes2021neural}, or the HpN synthesizer \cite{engel_ddsp_2020}.
Secondly, and most importantly, all DDSP resynthesis architectures require as conditioning the fundamental frequency from the target signal, extracted with an estimator \cite{kim_crepe_2018}. This ensures a harmonic alignment between the target and the prediction, and effectively avoids the problem of having to optimize pitch using gradient descent and the MSS loss during training.



A loss function that cannot propagate informative gradients cannot be employed to train a DNN. This is particularly disadvantageous for the case of FM generation, where the synthesis parameters include the modulation index $I$ and the frequency ratios $r$ that determine distance between the side-bands and the carrier. A small mismatch on ratio estimation can generate an unwanted vibrato-like effect, due to small frequency differences between oscillators. A big mismatch can hinder the joint optimization of the harmonics' positions and amplitudes. 
A DNN that does not learn how to precisely control the ratios could very easily incur either of those problems.






\section{Method}\label{sec:method}
\begin{figure*}
 \centering
 \includegraphics[width=1\textwidth]{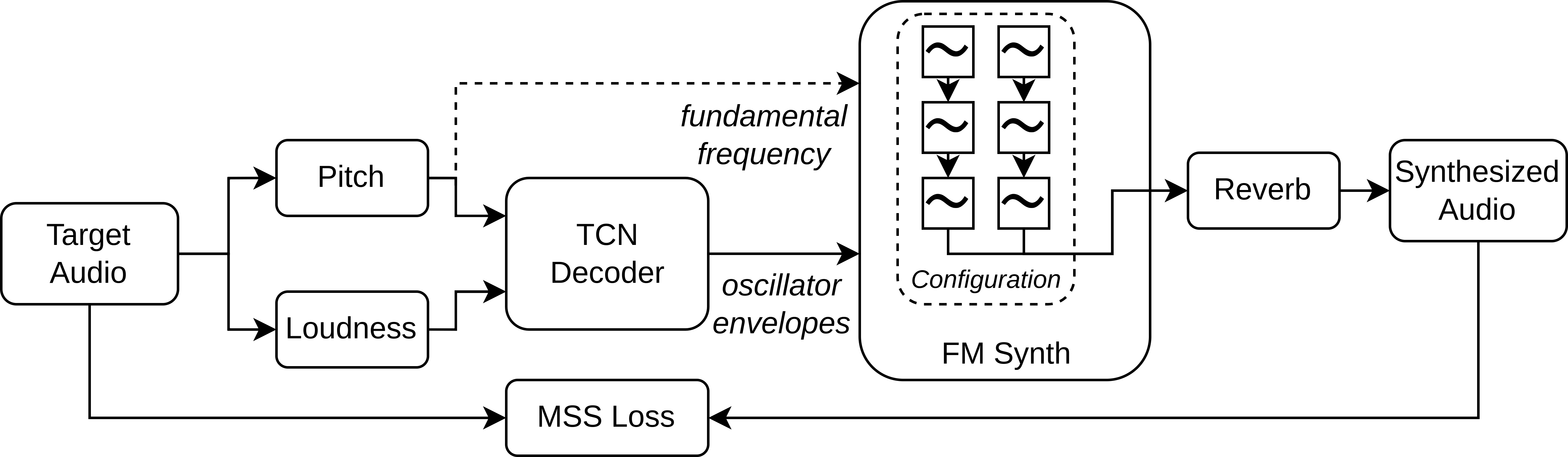}
 \caption{The DDX7 architecture employs a TCN decoder conditioned on a sequence of pitch and loudness frames to drive the envelopes of a few-oscillator differentiable FM synthesizer that features a fixed FM configuration with fixed frequency ratios, effectively mapping continuous controls of pitched musical instruments to a well-known synthesis architecture.}
 \label{fig:ddx7_architecture}
\end{figure*}

In this section, we propose a set of constraints that ease training of a DNN with a differentiable FM synthesizer. Next, we present DDX7, a NAS architecture for FM synthesis controlled by a Temporal Convolutional Network (TCN) \cite{bai_empirical_2018}. We train the DDX7 model for FM resynthesis of musical instrument sounds. This effectively results in a DX7 patch that is playable by an arbitrary audio input. A diagram of the architecture is shown in Figure \ref{fig:ddx7_architecture}.


\subsection{Differentiable DX7}

Our aim is to provide continuous control possibilities for a well-known FM synthesizer. We choose the Yamaha DX7 for its lasting influence on musical practice.


Considering the DX7 patch design, we observe that when fixing the frequency ratios and the routing of the oscillators, all the harmonics and overtones that can be generated take a fixed position in the spectrum.
This \textit{patch constraint} can allow the synthesis model to propagate informative gradients from the MSS reconstruction loss, provided the FM partials are pre-aligned with the ones of the target audio by means of pitch conditioning, as in the case of DDSP resynthesis.

\begin{figure}
 \centering\includegraphics[width=1\columnwidth]{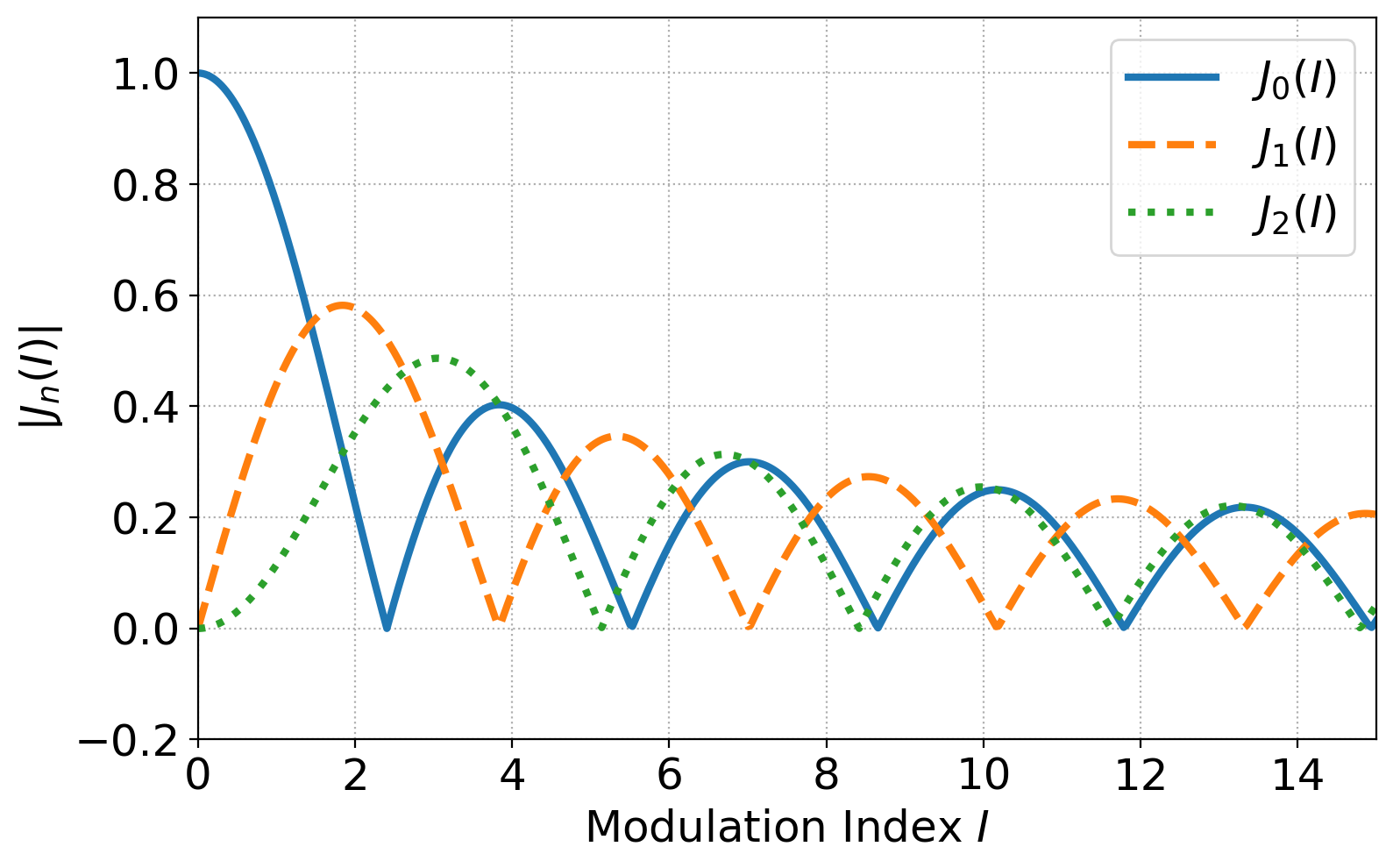}
 \caption{Absolute amplitudes of the first three harmonics generated by FM in function of $I$.}
 \label{fig:bessel_module}
\end{figure}

We propose a control scheme for an FM synthesizer where a DNN controls the modulation indices and volume of the oscillators. The oscillator routing and frequency ratios remain fixed. 
One problem may arise when considering the maximum values that the modulation index $I$ can take. For different ranges of $I$, the Bessel functions can create local minima during spectral optimization due to their oscillatory nature.
In the DX7, the modulation index envelopes can take values of as much as $4\pi$ \cite{fm_for_musicians_by_musicians}, but only for $I < 1.83$, are the Bessel functions strictly monotonic, with the carrier just exchanging energy with the side-bands, as shown in Figure \ref{fig:bessel_module}.
We analyze the effect of the maximum ranges through experiments in Section \ref{sec:evaluation}.

We implement the FM modulation algorithm in Pytorch, and adapt it manually for the different \textit{FM configurations} that are tried in this work. Each configuration defines a fixed oscillator routing and a set of frequency ratios.

\subsection{TCN Decoder}


TCNs have been successfully employed a number of sequential modelling tasks, including for audio processing and generation \cite{oord2016wavenet,michelashvili2020hierarchical,micro_tcn}. These are fully-convolutional networks that employ 1-dimensional convolutions and exponentially growing dilations to efficiently model long sequences within their receptive field \cite{bai_empirical_2018}.
We choose TCNs for our DDX7 implementation for their fast training capabilities and good sequential modelling performance.


In our FM resynthesis problem, we aim to map a set of synchronous input sequences of pitch ${f_0}_1, ... ,{f_0}_T \in \mathbb{R}$ and loudness $ld_1, ..., ld_T \in \mathbb{R}$ to the controls of our constrained synthesizer, i.e. the output levels $ol_1,...,ol_T \in \mathbb{R}^6$ of the six oscillators. 
We define the parameterized mapping function $f_{\theta}$ as shown in Equation \ref{eq:dx7decoder}, where the conditioning sequence $c_1,...,c_T \in \mathbb{R}^2$ is obtained by concatenating along a new dimension both pitch and loudness sequences, $\sigma(.)$ is the sigmoid activation function and $A_{max}$ is the maximum output level value that the envelopes can take for that oscillator, as shown in Equation \ref{eq:a_max}, where $I_{max}$ is a hyperparameter describing the maximum modulation index range that the system can realize.

\begin{equation}
    \hat{ol_1},...\hat{ol_T} = A_{max} \cdot \sigma (f_{\theta}(c_1,...c_T))
\label{eq:dx7decoder}
\end{equation}
\begin{equation}
    A_{max} = 
\begin{cases}
    1       & \text{if carrier }\\
    I_{max} & \text{otherwise}
\end{cases}
\label{eq:a_max}
\end{equation}

We implement our mapping function $f_{\theta}$ with a simple, causal TCN architecture following \cite{bai_empirical_2018}, with 2 input and 6 output channels, processed by 5 TCN residual blocks with skip connections and 128 hidden channels each.
Each residual block features two convolutional layers of kernel size 3, and the dilation increases by a factor of 2 in each block. 
Weight normalization, dropout with probability of 0.5 and ReLU activation functions are used throughout the network, with the exception of the output layer that features a sigmoid layer.
This yields a relatively lightweight decoder, with about 400k parameters, and a receptive field $T$ of 125 pitch and loudness frames.

\subsection{Learnable Reverb}

We employ a differentiable reverb module, similar to the one employed for the DDSP decoder \cite{engel_ddsp_2020}, featuring learnable mix and decay parameters, and a trainable impulse response of 1 second length. This is used to estimate the room response of the dataset recordings, decoupling it from the FM sound generation block. It is applied directly to the FM synthesizer output and it is jointly optimized with the DNN during training.

\section{Training}\label{sec:training}

\subsection{Loss function}


The DDX7 architecture can be trained with a corpus of audio from a musical instrument as supervision, employing stochastic gradient descent on minibatches with a spectral reconstruction objective.
We employ the MSS reconstruction loss shown in Equation \ref{eq:rec_loss}, where $S_i$ and $\hat{S}_i$ are the magnitude spectrograms of the target and synthesized audio respectively, $||.||_1$ denotes the $L_1$ norm, and $i$ is a particular Fourier transform analysis window on which the spectrogram is computed. We use $i \in \{64,128,256,512,1024,2048\} $ with an overlap of 75\% between windows.
\begin{equation}
    L = \sum_i ( ||S_i - \hat{S}_i||_1 + || log S_i - log \hat{S_i}||_1 )
\label{eq:rec_loss}\end{equation}

\subsection{Dataset}

We train our DDX7 models on audio samples of instruments extracted from the University of Rochester Music Performance (URMP) dataset \cite{li_creating_2019}, an audio-visual dataset that contains classical pieces.
We select the separated audio stems for violin, flute and trumpet as our training data, down-sample them to 16 kHz, remove silences and crop the audio files to instances of 4 seconds. We extract the A-weighted loudness \cite{moore1997model} and fundamental frequency employing the CREPE \cite{kim_crepe_2018} pitch estimator. We discard all instances that yield a mean pitch confidence smaller than 0.85, with the exception of the flute corpus, for which we relax the requirement down to 0.80 due to its short length. We further normalize pitch and loudness values within a range between 0 and 1.

We process each annotation with a hop size of 64 samples, yielding pitch and loudness sequences of 1000 frames for each 4-second instance. The selection of hop size and sample rate results in our TCN model featuring a receptive field of 0.5 s, and dictates the frame rate at which it drives the oscillators, 250 Hz. We linearly interpolate the envelope frames before feeding them into the oscillators. Finally, we separate the dataset into train, validation and test sets with 0.75 / 0.125 / 0.125 splits respectively.


\subsection{FM configurations}

\begin{figure}
 \centering
 \includegraphics[width=1\columnwidth]{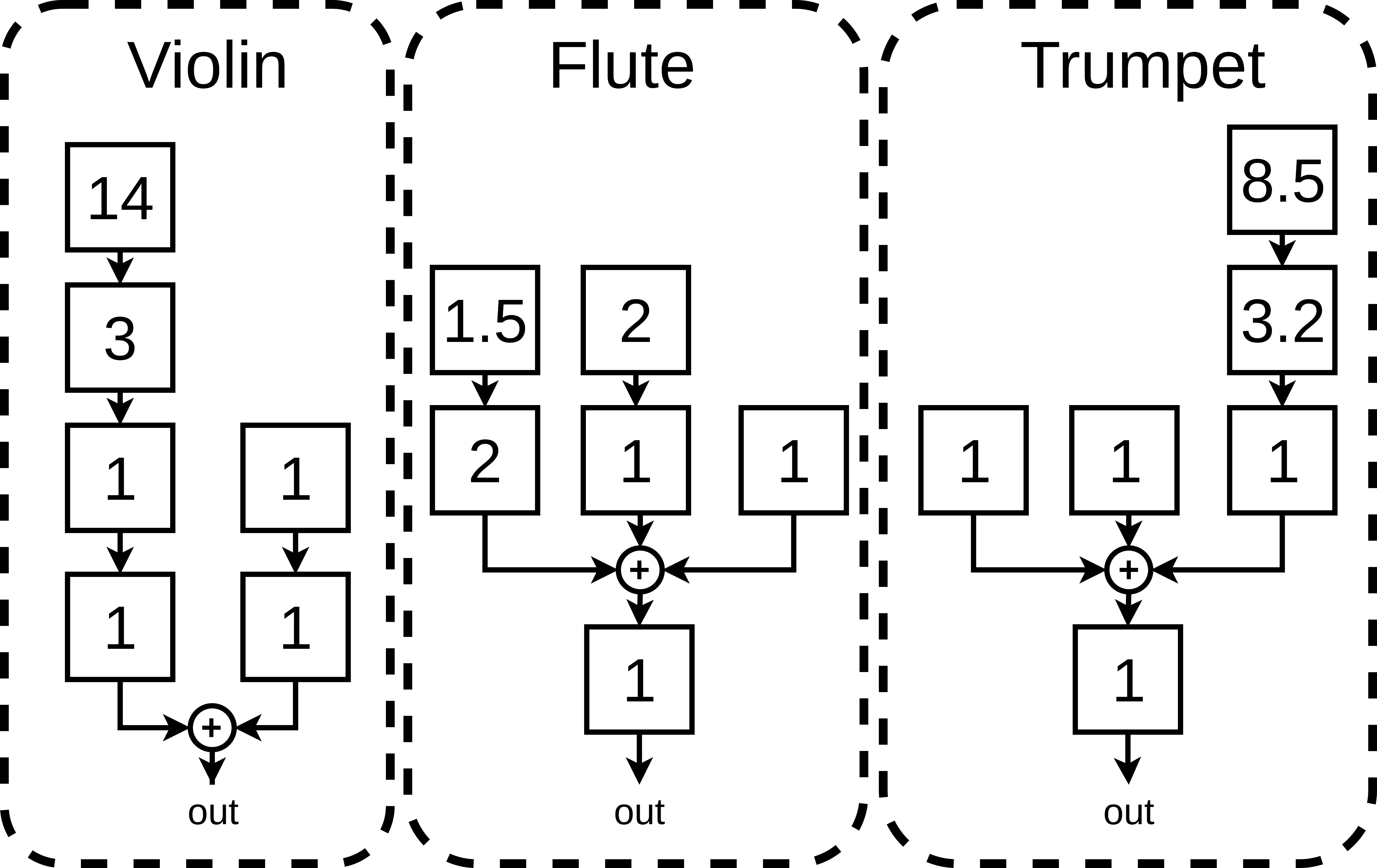}
 \caption{FM configurations used for training. Squares indicate sinusoidal oscillators and their frequency ratios.}
 \label{fig:patches}
\end{figure}

For each target instrument, we select a different FM configuration extracted from the original patch set of the Yamaha DX7, which we retrieve from the web.\footnote{\url{http://bobbyblues.recup.ch/yamaha_dx7/}} Then, we load the patches in Dexed \cite{dexed_repo}, a DX7 emulator, and audit them searching for most similar to the target instruments. We select "STRINGS 1" for violin, "FLUTE 1" for flute and "BRASS 3" for the trumpet. We deploy the oscillator routing with the frequency ratios rounded up to one decimal point (to avoid vibrato-like effects) as differentiable FM modules, as shown in Figure \ref{fig:patches}. We do not deploy the oscillator feedback feature of the DX7 in our implementation, as it cannot be computed in parallel and we find it is very slow to render using a standard for loop in Python.

\subsection{Training process}

Our models are trained for 120k steps, with the Adam optimizer, set with an initial learning rate of 3e-4, decreasing with a factor of 0.98 each 10k steps. We clip gradients to a maximum norm of 2, and employ a batch size of 16.




\section{Evaluation}\label{sec:evaluation}
We evaluate the performance of DDX7 on the resynthesis task, training the model on the flute, trumpet and violin corpus, and evaluate its performance on selected benchmarks with the Fréchet Audio Distance (FAD).

\subsection{Fréchet audio distance}

The Fréchet Audio Distance (FAD) presented by Kilgour et. al. \cite{kilgour_frechet_2019} serves as a quality metric for audio enhancement, that correlates better with human listeners than SDR (signal-to-distortion ratio) or spectral differences such as the MSS loss. This is in line with other deep neural features used in Computer Vision that are found to outperform heuristic metrics by great margins \cite{zhang_unreasonable_2018}. It has been used to assess synthesis quality in previous works \cite{hayes2021neural,nistal_drumgan_2020,lavault2022stylewavegan}.
The FAD computes the Fréchet distance between multivariate Gaussian distributions inferred from the embeddings of a pre-trained VGGish model\cite{hershey_cnn_2017}.
The compared distributions are generated from the embeddings of a corpus of audio for evaluation and a "background" corpus of high quality audio as reference.

\subsection{Benchmarks}

\subsubsection{Maximum modulation index}

We are interested into knowing which are the best modulation index limits for which our model can successfully render the instrument audio. Taking into account that optimizing for a particular spectra may be difficult for the original maximum modulation index range of the DX7 of $4\pi$, we train our model for each instrument with three different maximum modulation index ranges for the oscillators: $I_{max} = \{4\pi,2\pi, 2\}$, including the original DX7 range, a halved one and one limited at two, which includes the global maximum of the first harmonic $J_1(I)$, but limits the other Bessel functions to a space where they are strictly monotonic.
Finally, we load the corresponding configuration for each instrument selected, and we train three versions of DDX7, to account for each of the values of $I_{max}$. 

Our FM model learns to control the envelopes of 6 oscillators frame-wise. As a baseline, we assess the performance of the common Harmonic plus Noise spectral modelling synthesizer employed for resynthesis tasks. We train a baseline pitch and loudness decoder that controls 121 frame-wise parameters of an HpN synthesizer, similar to the original solo instrument DDSP Decoder \cite{engel_ddsp_2020}, but implemented in the same training framework as DDX7, to ensure that we use exactly the same dataset, preprocessing and frame rate. The model is roughly 11 times bigger than DDX7, with about 4.5 M parameters. We train it for 120k steps for parity with our model, with a batch size of 16, and an initial learning rate of 1e-4, decreasing to a factor of 0.98 for each 10k steps.

For evaluation with FAD, we generate background embedding distributions of the complete audio corpus of each instrument. Next, we generate embeddings distributions from the resynthesized excerpts of the test set for each model and instrument. We compute the FAD of these and the original test set against their corresponding background distribution.
Results are shown in Table \ref{tab:max_ol}. We observe that the HpN baseline outperforms our model in violin and trumpet, but controlling 121 parameters instead of 6, and at a higher computational cost. Surprisingly, DDX7 outperforms the baseline on flute, suggesting that our \textit{patch constraint} approach can bias a DNN towards a usable FM timbral space.
We observe the best DDX7 performance is obtained with $I_{max} = 2$ for flute and violin and $I_{max} = 2\pi$ for trumpet. This may be correlated with the different spectra of the instruments, suggesting that the trumpet may require a bigger $I_{max}$ for an improved reconstruction. 
Finally, for some configurations of $I_{max}$ the models sound unnatural and fail at the estimation of the room response. We leave further analysis of this issue for future work.

\begin{figure}
 \centering
 \includegraphics[width=1\columnwidth]{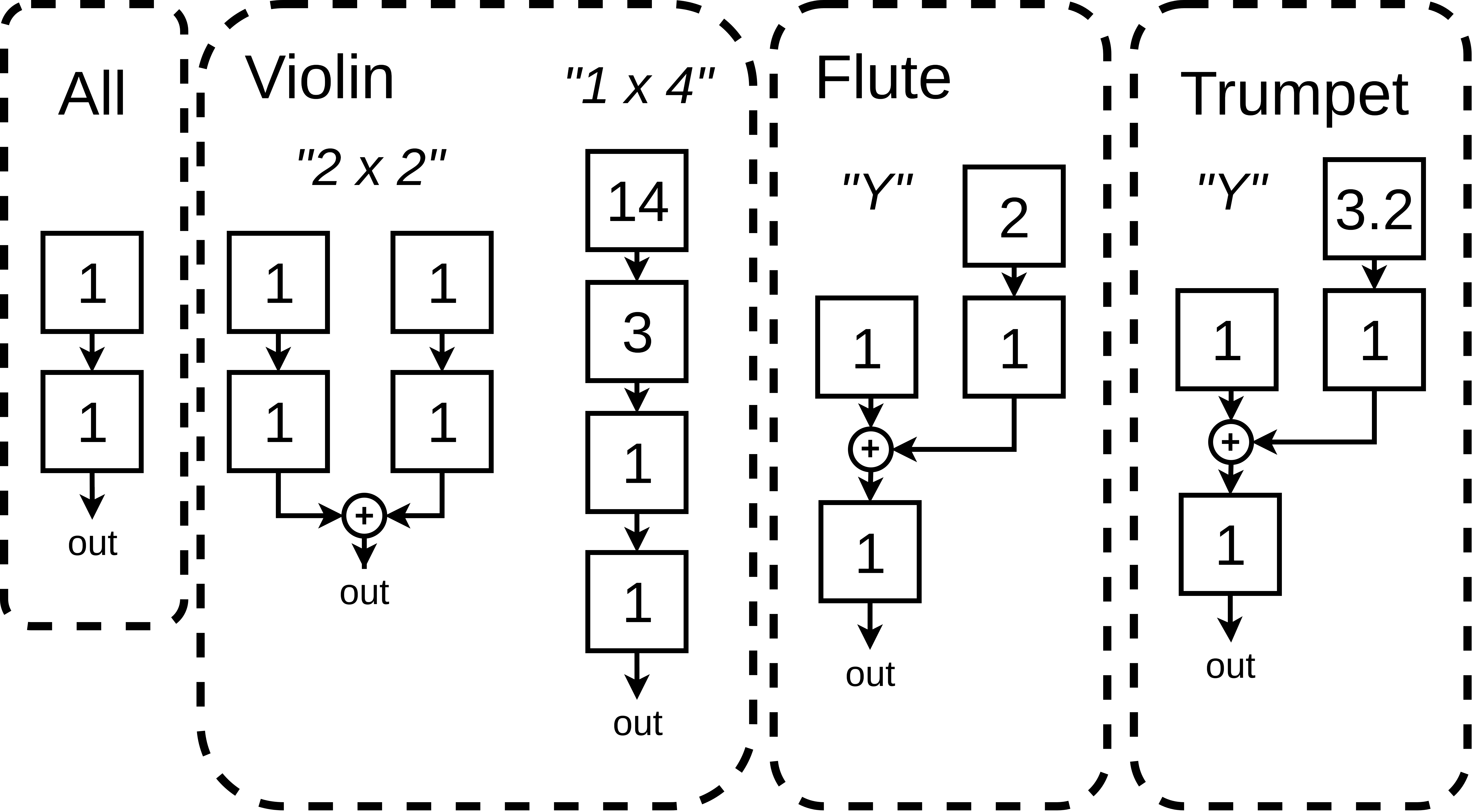}
 \caption{Ablated FM configurations.}
 \label{fig:ablated_patches}
\end{figure}

\begin{table}
\centering
\begin{tabular}{|l|ccc|}
\hline
               & \multicolumn{3}{c|}{\textbf{Fréchet Audio Distance}}                                          \\ \hline
\textbf{Model} & \multicolumn{1}{c|}{\textit{Flute}} & \multicolumn{1}{c|}{\textit{Violin}} & \textit{Trumpet} \\ \hline
Test Data      & \multicolumn{1}{c|}{2.074}          & \multicolumn{1}{c|}{0.577}           & 1.069            \\ \hline
HpN Baseline   & \multicolumn{1}{c|}{4.326}          & \multicolumn{1}{c|}{\textbf{0.795}}           & \textbf{2.486}            \\ \hline
DDX7 ($I_{max}=2$)       & \multicolumn{1}{c|}{\underline{\textbf{2.731}}}          & \multicolumn{1}{c|}{\underline{1.618}}           & 4.941            \\
DDX7 ($I_{max}=2\pi$)    & \multicolumn{1}{c|}{3.281}          & \multicolumn{1}{c|}{2.148}           & \underline{3.326}            \\
DDX7 ($I_{max}=4\pi$)    & \multicolumn{1}{c|}{2.938}          & \multicolumn{1}{c|}{1.637}           & 3.853            \\ \hline
\end{tabular}
\caption{
FAD of resynthesis results for all models computed against the background embedding distributions for each instrument complete corpus. Best results are in bold and best $I_{max}$ configurations are underlined.}
 \label{tab:max_ol}
\end{table}

\subsubsection{Oscillator ablation test}

We want to assess if our model is effectively optimized to leverage all of the modulators for the reconstruction task. We propose ablated versions of the previous patches with two and four oscillators, as shown in Figure \ref{fig:ablated_patches}.
We train the DDX7 models on the ablated patches using the optimal $I_{max}$ found with the previous benchmark and compare their resynthesis quality with the FAD following the same previous procedure.
The results shown in Table \ref{tab:abl_test} suggest that the violin and flute model benefit from the extra degrees of freedom present with more oscillators. On the other hand, the trumpet model works best with the smallest configuration, possibly due to an incorrect patch selection that hindered the optimization process. Finally, the 2-oscillator trumpet model outperforms the HpN baseline, suggesting that good results can be achieved with a small number of frequency-modulated oscillators.


\begin{table}
\begin{tabular}{|l|ccccc|}
\hline
\multirow{2}{*}{\textbf{Instrument}} & \multicolumn{5}{c|}{\textbf{FM Configuration}}                                                                                                         \\ \cline{2-6} 
                                     & \multicolumn{1}{c|}{\textit{6}} & \multicolumn{1}{c|}{\textit{4 "Y"}} & \multicolumn{1}{c|}{\textit{4x1}} & \multicolumn{1}{c|}{\textit{2x2}} & \textit{2}     \\ \hline
\textit{Flute}                                & \textbf{2.731}                  & 3.246                               & -                                 & -                                 & 3.364          \\ \hline
\textit{Violin}                               & \textbf{1.618}                  & -                                   & 1.877                             & 5.620                             & 8.270          \\ \hline
\textit{Trumpet}                              & 3.326                           & 2.943                               & -                                 & -                                 & \textbf{1.674} \\ \hline
\end{tabular}
\caption{
FAD for complete and ablated patches on DDX7.}
 \label{tab:abl_test}
\end{table}








\section{Conclusion}\label{sec:conclusion}

We presented DDX7, an approach for FM resynthesis of musical instrument sounds that yields good reconstructions controlling few parameters, with relatively smaller models. We have shown that FM with a \textit{patch constraint} can perform comparably well to a more complex baseline with just 6, and even less oscillators; we hope this motivates further research along this line, including for instance sound matching techniques to find suitable configurations.

Current resynthesis architectures feature synthesizers that are difficult to intervene in a musically meaningful way. In contrast, DDX7 learns to control an FM synthesizer that is common in the sound design practice. It replaces the ADSR generator of the original DX7 with a TCN that infers the envelopes from continuous control inputs. At runtime, it is possible to manipulate the timbre on-the-fly, either by re-shaping the spectrum with the ratios, altering dynamics on the envelopes, or by re-routing the oscillators. Finally, the small model size and causal temporal dependency make DDX7 an interesting candidate for real-time implementation.
We leave an exploration of these affordances and possibilities for future work.





\section{Acknowledgements}

We would like to thank the ISMIR reviewers for their valuable feedback. Also, we would like to thank our colleagues Ben Hayes and Rodrigo Diaz for their advice and many compelling discussions about audio rendering and DNN optimization. This work was supported by UK Research and Innovation [grant number EP/S022694/1]. AM's contributions are supported by the Royal Academy of Engineering under the Research Chairs and Senior Research Fellowships scheme.

\bibliography{ISMIR2022_template}

\begin{thebibliography}{10}
\providecommand{\url}[1]{#1}
\csname url@samestyle\endcsname
\providecommand{\newblock}{\relax}
\providecommand{\bibinfo}[2]{#2}
\providecommand{\BIBentrySTDinterwordspacing}{\spaceskip=0pt\relax}
\providecommand{\BIBentryALTinterwordstretchfactor}{4}
\providecommand{\BIBentryALTinterwordspacing}{\spaceskip=\fontdimen2\font plus
\BIBentryALTinterwordstretchfactor\fontdimen3\font minus
  \fontdimen4\font\relax}
\providecommand{\BIBforeignlanguage}[2]{{%
\expandafter\ifx\csname l@#1\endcsname\relax
\typeout{** WARNING: IEEEtran.bst: No hyphenation pattern has been}%
\typeout{** loaded for the language `#1'. Using the pattern for}%
\typeout{** the default language instead.}%
\else
\language=\csname l@#1\endcsname
\fi
#2}}
\providecommand{\BIBdecl}{\relax}
\BIBdecl

\bibitem{chowning1973synthesis}
J.~M. Chowning, ``The synthesis of complex audio spectra by means of frequency
  modulation,'' \emph{Journal of the Audio Engineering Society}, vol.~21,
  no.~7, pp. 526--534, 1973.

\bibitem{lavengood2019makes}
M.~Lavengood, ``What makes it sound ’80s? {T}he {Yamaha DX7} electric piano
  sound,'' \emph{Journal of Popular Music Studies}, vol.~31, no.~3, pp. 73--94,
  2019.

\bibitem{miranda2012computer}
E.~Miranda, \emph{{Computer Sound Design: Synthesis techniques and
  programming}}.\hskip 1em plus 0.5em minus 0.4em\relax Routledge, 2012.

\bibitem{stevens2021teaching}
B.~Stevens, \emph{Teaching Electronic Music: Cultural, Creative, and Analytical
  Perspectives}.\hskip 1em plus 0.5em minus 0.4em\relax Routledge, 2021.

\bibitem{pinch1998social}
T.~Pinch and F.~Trocco, ``The social construction of the early electronic music
  synthesizer,'' \emph{ICON}, pp. 9--31, 1998.

\bibitem{the_nime_reader}
A.~R. Jensenius and M.~J. Lyons, \emph{A NIME reader: Fifteen years of New
  Interfaces for Musical Expression}.\hskip 1em plus 0.5em minus 0.4em\relax
  Springer, 2017, vol.~3.

\bibitem{west2021making}
T.~West, B.~Caramiaux, S.~Huot, and M.~M. Wanderley, ``Making mappings: Design
  criteria for live performance,'' \emph{New Interfaces for Musical Expression
  conference (NIME)}, 5 2021.

\bibitem{regimbal2021interpolating}
J.~Regimbal and M.~M. Wanderley, ``Interpolating audio and haptic control
  spaces,'' in \emph{New Interfaces for Musical Expression conference
  (NIME)}.\hskip 1em plus 0.5em minus 0.4em\relax PubPub, 2021.

\bibitem{asdss_paper}
C.~Poepel and R.~B. Dannenberg, ``{Audio Signal Driven Sound Synthesis},'' in
  \emph{International Computer Music Conference}, 2005.

\bibitem{1678000}
V.~Verfaille, U.~Zolzer, and D.~Arfib, ``Adaptive digital audio effects
  (a-dafx): a new class of sound transformations,'' \emph{IEEE Transactions on
  Audio, Speech, and Language Processing}, vol.~14, no.~5, pp. 1817--1831,
  2006.

\bibitem{nuimeprn4166}
V.~Lazzarini, J.~Timoney, and T.~Lysaght, ``Adaptive {FM} {S}ynthesis,'' in
  \emph{DAFX-07 the 10th Int. Conference on Digital Audio Effects}, September
  2007.

\bibitem{engel_ddsp_2020}
J.~Engel, L.~H. Hantrakul, C.~Gu, and A.~Roberts, ``{DDSP}: {Differentiable}
  {Digital} {Signal} {Processing},'' in \emph{8th {International} {Conference}
  on {Learning} {Representations}}, Addis Ababa, Ethiopia, 2020.

\bibitem{hayes2021neural}
B.~Hayes, C.~Saitis, and G.~Fazekas, ``Neural waveshaping synthesis,''
  \emph{Proceedings of the 22th International Society for Music Information
  Retrieval Conference}, 2021.

\bibitem{michelashvili2020hierarchical}
M.~Michelashvili and L.~Wolf, ``Hierarchical timbre-painting and articulation
  generation,'' \emph{Proceedings of the 21th International Society for Music
  Information Retrieval Conference}, 2020.

\bibitem{cifka2021self}
O.~Cifka, A.~Ozerov, U.~{\c{S}}im{\c{s}}ekli, and G.~Richard,
  ``Self-{S}upervised {VQ-VAE} for {O}ne-{S}hot {M}usic {S}tyle {T}ransfer,''
  in \emph{ICASSP 2021-2021 IEEE International Conference on Acoustics, Speech
  and Signal Processing (ICASSP)}.\hskip 1em plus 0.5em minus 0.4em\relax IEEE,
  2021, pp. 96--100.

\bibitem{50554}
M.~Carney, C.~Li, E.~Toh, P.~Yu, and J.~Engel, ``Tone transfer: In-browser
  interactive neural audio synthesis,'' in \emph{Joint Proceedings of the ACM
  IUI 2021 Workshops}, 2021.

\bibitem{wu2021midi}
Y.~Wu, E.~Manilow, Y.~Deng, R.~Swavely, K.~Kastner, T.~Cooijmans, A.~Courville,
  C.-Z.~A. Huang, and J.~Engel, ``{MIDI-DDSP}: Detailed control of musical
  performance via hierarchical modeling,'' \emph{International Conference on
  Learning Representations (ICLR) 2022}, 2022.

\bibitem{yee2021studio}
M.~Yee-King and L.~McCallum, ``Studio report: Sound synthesis with {DDSP} and
  network bending techniques,'' \emph{Proceedings of the 2nd Conference on AI
  Music Creativity}, 2021.

\bibitem{nielsen2020practical}
K.~Nielsen, ``Practical linear and exponential frequency modulation for digital
  music synthesis,'' \emph{Proceedings of the 23rd International Conference on
  Digital Audio Effects (DAFx-20), Vienna, Austria, September 8–12, 2020-21},
  2020.

\bibitem{masuda2021quality}
N.~Masuda and D.~Saito, ``Quality diversity for synthesizer sound matching,''
  in \emph{Proceedings of the 23rd International Conference on Digital Audio
  Effects (DAFx20in21)}, 2021.

\bibitem{fm_devices_1}
\BIBentryALTinterwordspacing
{Sound On Sound Magazine}. (2020) {Korg Opsix}. [Online]. Available:
  \url{https://www.soundonsound.com/reviews/korg-opsix}
\BIBentrySTDinterwordspacing

\bibitem{fm_devices_2}
\BIBentryALTinterwordspacing
{Max Kuehn, for Fidlar Music}. (2022) {Best {FM} Synth 2022}. [Online].
  Available: \url{https://fidlarmusic.com/best-fm-synth/}
\BIBentrySTDinterwordspacing

\bibitem{horner1993machine}
A.~Horner, J.~Beauchamp, and L.~Haken, ``Machine tongues {XVI}: Genetic
  algorithms and their application to {FM} matching synthesis,'' \emph{Computer
  Music Journal}, vol.~17, no.~4, pp. 17--29, 1993.

\bibitem{yee-king_automatic_2018}
M.~J. Yee-King, L.~Fedden, and M.~d'Inverno,
  ``\BIBforeignlanguage{en}{Automatic {Programming} of {VST} {Sound}
  {Synthesizers} {Using} {Deep} {Networks} and {Other} {Techniques}},''
  \emph{\BIBforeignlanguage{en}{IEEE Transactions on Emerging Topics in
  Computational Intelligence}}, vol.~2, no.~2, pp. 150--159, Apr. 2018.

\bibitem{le2021improving}
G.~Le~Vaillant, T.~Dutoit, and S.~Dekeyser, ``Improving synthesizer programming
  from variational autoencoders latent space,'' in \emph{Proceedings of the
  23rd International Conference on Digital Audio Effects (DAFx20in21)}, 2021.

\bibitem{chen2022sound2synth}
Z.~Chen, Y.~Jing, S.~Yuan, Y.~Xu, J.~Wu, and H.~Zhao, ``{Sound2Synth}:
  Interpreting sound via {FM} synthesizer parameters estimation,'' \emph{arXiv
  preprint arXiv:2205.03043}, 2022.

\bibitem{oord2016wavenet}
A.~v.~d. Oord, S.~Dieleman, H.~Zen, K.~Simonyan, O.~Vinyals, A.~Graves,
  N.~Kalchbrenner, A.~Senior, and K.~Kavukcuoglu, ``Wavenet: A generative model
  for raw audio,'' \emph{The 9th ISCA Speech Synthesis Workshop}, 2016.

\bibitem{sample_rnn}
S.~Mehri, K.~Kumar, I.~Gulrajani, R.~Kumar, S.~Jain, J.~Sotelo, A.~Courville,
  and Y.~Bengio, ``\BIBforeignlanguage{en}{{SampleRNN}: {An} {Unconditional}
  {End}-to-{End} {Neural} {Audio} {Generation} {Model}},'' in
  \emph{\BIBforeignlanguage{en}{5th {International} {Conference} on {Learning}
  {Representations}}}, Toulon, France, 2017.

\bibitem{nistal_drumgan_2020}
J.~Nistal, S.~Lattner, and G.~Richard, ``{DrumGAN}: {Synthesis} of {Drum}
  {Sounds} {With} {Timbral} {Feature} {Conditioning} {Using} {Generative}
  {Adversarial} {Networks},'' in \emph{Proceedings of the 21th {International}
  {Society} for {Music} {Information} {Retrieval} {Conference}}, Montréal,
  Aug. 2020.

\bibitem{engel_gansynth_2019}
J.~Engel, K.~K. Agrawal, S.~Chen, I.~Gulrajani, C.~Donahue, and A.~Roberts,
  ``\BIBforeignlanguage{en}{{GANSynth}: {Adversarial} {Neural} {Audio}
  {Synthesis}},'' in \emph{\BIBforeignlanguage{en}{7th {International}
  {Conference} on {Learning} {Representations}}}, New Orleans, LA, USA, 2019,
  p.~17.

\bibitem{lavault2022stylewavegan}
A.~Lavault, A.~Roebel, and M.~Voiry, ``{StyleWaveGAN: Style-based synthesis of
  drum sounds with extensive controls using generative adversarial networks},''
  \emph{arXiv preprint arXiv:2204.00907}, 2022.

\bibitem{esling_flow_2020}
P.~Esling, N.~Masuda, A.~Bardet, R.~Despres, and A.~Chemla-Romeu-Santos,
  ``\BIBforeignlanguage{en}{Flow {Synthesizer}: {Universal} {Audio}
  {Synthesizer} {Control} with {Normalizing} {Flows}},''
  \emph{\BIBforeignlanguage{en}{Applied Sciences}}, vol.~10, no.~1, p. 302,
  2020.

\bibitem{caillon2021rave}
A.~Caillon and P.~Esling, ``{RAVE: A variational autoencoder for fast and
  high-quality neural audio synthesis},'' \emph{arXiv preprint
  arXiv:2111.05011}, 2021.

\bibitem{huang_timbretron_2019}
S.~Huang, Q.~Li, C.~Anil, S.~Oore, and R.~B. Grosse,
  ``\BIBforeignlanguage{en}{{TimbreTron} {A}
  {WaveNet}({CycleGAN}({CQT}({Audio}))) {Pipeline} for {Musical} {Timbre}
  {Transfer}},'' in \emph{\BIBforeignlanguage{en}{7th {International}
  {Conference} on {Learning} {Representations}}}, New Orleans, LA, USA, 2019,
  p.~17.

\bibitem{wang2019neural}
X.~Wang, S.~Takaki, and J.~Yamagishi, ``Neural source-filter waveform models
  for statistical parametric speech synthesis,'' \emph{IEEE/ACM Transactions on
  Audio, Speech, and Language Processing}, vol.~28, pp. 402--415, 2019.

\bibitem{stylianou1995high}
Y.~Stylianou, J.~Laroche, and E.~Moulines, ``High-quality speech modification
  based on a harmonic+ noise model,'' in \emph{Fourth European Conference on
  Speech Communication and Technology}, 1995.

\bibitem{andreas_jansson_repo}
A.~Jansson, ``Implicit neural differentiable {FM} synthesizer,''
  \url{https://github.com/andreasjansson/fmsynth}, 2022.

\bibitem{juan_alonso_repo}
J.~Alonso, ``{DDSP-FM}: differentiable {FM} synthesis,''
  \url{https://juanalonso.github.io/ddsp_fm/}, 2021.

\bibitem{turian_im_2020}
J.~Turian and M.~Henry, ``I'm {Sorry} for {Your} {Loss}: {Spectrally}-{Based}
  {Audio} {Distances} {Are} {Bad} at {Pitch},'' \emph{arXiv:2012.04572 [cs,
  eess]}, Dec. 2020, {I} Can't Believe It's Not Better! (ICBINB) NeurIPS 2020
  Workshop.

\bibitem{kim_crepe_2018}
J.~W. Kim, J.~Salamon, P.~Li, and J.~P. Bello, ``\BIBforeignlanguage{English
  (US)}{Crepe: {A} {Convolutional} {Representation} for {Pitch}
  {Estimation}},'' in \emph{\BIBforeignlanguage{English (US)}{2018 {IEEE}
  {International} {Conference} on {Acoustics}, {Speech}, and {Signal}
  {Processing}, {ICASSP} 2018 - {Proceedings}}}.\hskip 1em plus 0.5em minus
  0.4em\relax Institute of Electrical and Electronics Engineers Inc., Sep.
  2018, pp. 161--165.

\bibitem{bai_empirical_2018}
S.~Bai, J.~Z. Kolter, and V.~Koltun, ``An {Empirical} {Evaluation} of {Generic}
  {Convolutional} and {Recurrent} {Networks} for {Sequence} {Modeling},''
  \emph{arXiv:1803.01271 [cs]}, Apr. 2018, arXiv: 1803.01271.

\bibitem{fm_for_musicians_by_musicians}
D.~Bristow and J.~Chowning, ``{FM Theory and Applications: By Musicians for
  Musicians},'' \emph{Yamaha Music Foundation}, 1986.

\bibitem{micro_tcn}
C.~J. Steinmetz and J.~D. Reiss, ``Efficient neural networks for real-time
  modeling of analog dynamic range compression,'' in \emph{152nd AES
  Convention}, 2022.

\bibitem{li_creating_2019}
B.~Li, X.~Liu, K.~Dinesh, Z.~Duan, and G.~Sharma, ``Creating a {Multitrack}
  {Classical} {Music} {Performance} {Dataset} for {Multimodal} {Music}
  {Analysis}: {Challenges}, {Insights}, and {Applications},'' \emph{IEEE
  Transactions on Multimedia}, vol.~21, no.~2, pp. 522--535, Feb. 2019.

\bibitem{moore1997model}
B.~C. Moore, B.~R. Glasberg, and T.~Baer, ``A model for the prediction of
  thresholds, loudness, and partial loudness,'' \emph{Journal of the Audio
  Engineering Society}, vol.~45, no.~4, pp. 224--240, 1997.

\bibitem{dexed_repo}
P.~Gauthier, ``Dexed - {FM} {P}lugin {S}ynth,''
  \url{https://github.com/asb2m10/dexed}, 2022.

\bibitem{kilgour_frechet_2019}
K.~Kilgour, M.~Zuluaga, D.~Roblek, and M.~Sharifi,
  ``\BIBforeignlanguage{en}{Fréchet {Audio} {Distance}: {A} {Reference}-{Free}
  {Metric} for {Evaluating} {Music} {Enhancement} {Algorithms}},'' in
  \emph{\BIBforeignlanguage{en}{Interspeech 2019}}.\hskip 1em plus 0.5em minus
  0.4em\relax ISCA, Sep. 2019, pp. 2350--2354.

\bibitem{zhang_unreasonable_2018}
R.~Zhang, P.~Isola, A.~A. Efros, E.~Shechtman, and O.~Wang,
  ``\BIBforeignlanguage{en}{The {Unreasonable} {Effectiveness} of {Deep}
  {Features} as a {Perceptual} {Metric}},'' in
  \emph{\BIBforeignlanguage{en}{2018 {IEEE}/{CVF} {Conference} on {Computer}
  {Vision} and {Pattern} {Recognition}}}.\hskip 1em plus 0.5em minus
  0.4em\relax Salt Lake City, UT: IEEE, Jun. 2018, pp. 586--595.

\bibitem{hershey_cnn_2017}
S.~Hershey, S.~Chaudhuri, D.~P.~W. Ellis, J.~F. Gemmeke, A.~Jansen, R.~C.
  Moore, M.~Plakal, D.~Platt, R.~A. Saurous, B.~Seybold, M.~Slaney, R.~J.
  Weiss, and K.~Wilson, ``{CNN} architectures for large-scale audio
  classification,'' in \emph{2017 {IEEE} {International} {Conference} on
  {Acoustics}, {Speech} and {Signal} {Processing} ({ICASSP})}.\hskip 1em plus
  0.5em minus 0.4em\relax New Orleans, LA: IEEE, Mar. 2017, pp. 131--135.

\end{thebibliography}

%
%
%
%
%

\end{document}